\documentclass[aps, prd, twocolumn, superscriptaddress, nofootinbib, preprintnumbers]{revtex4}
\usepackage[dvipdfmx,hiresbb]{graphicx} 
\usepackage{graphicx}
\usepackage{amsmath,amssymb}
\usepackage{ulem} 
\usepackage{epsf}
\usepackage{color}   
\usepackage{here}
\allowdisplaybreaks[1]

\newcommand{\nn}{\nonumber\\}

\newcommand{\be}{\begin{equation}}
\newcommand{\e}{\end{equation}}
\newcommand{\aln}[1]{\begin{align}#1\end{align}}
\newcommand{\eqn}[1]{\begin{eqnarray}#1\end{eqnarray}}

\begin{document}
\preprint{
KEK-TH 2045
}
\title{
Amplification of gravitational motion via Quantum weak measurement   
}  
\date{\today}

\author{Kiyoharu Kawana}
\affiliation{
KEK Theory Center, IPNS, Ibaraki 305-0801, Japan
}

\author{Daiki Ueda}
\affiliation{
KEK Theory Center, IPNS, Ibaraki 305-0801, Japan
}
\affiliation{
The Graduate University of Advanced Studies (Sokendai), Tsukuba, Ibaraki 305-0801, Japan
}

\begin{abstract}
We investigate a new experimental possibility of measuring the Newtonian gravitational constant $G$ by using the weak measurement. 
Amplification via weak measurement is one of the interesting phenomena of quantum mechanics. 
In this letter, we consider it in a system consisting of many cold atoms which are gravitationally interacting with an external macroscopic source and show that it is possible to obtain ${\cal{O}}(10^{3})$ amplification of their relative motion compared with the classical motion when the number of atoms are ${\cal{O}}(10^{15})$ and the observing time is $\sim 0.5$s. 
%
This result suggests that 
it might be possible to use this system as a new experimental set up for determining $G$.  
%
Besides, our study indicates that the gravitational force can behave as a repulsive force because of the  weak measurement.  
\end{abstract}

\maketitle
{\it Introduction.---}
Over the last few decades, various fundamental physical constants such as the Planck constant $\hbar$, the Avogadro constant $N_A^{}$, the fine structure constant $\alpha$ have been determined precisely, whose standard relative uncertainties are typically ${\cal{O}}(10^{-10\sim -8})$ \cite{Mohr:2015ccw}. 
On the other hand, the measurement of the Newtonian gravitational constant $G$ is not so precise as those physical constants, and its current standard relative uncertainty is ${\cal{O}}(10^{-4})$. 
There are mainly two reasons for this inaccuracy: 
%
One is of course due to the weakness of $G$. 
As long as we consider an experiment whose typical energy scale (or length $l$) is $p\ (l^{-1})\ll \sim G^{-1/2}$, the effect of gravity is too small ${\cal{O}}(p^2G)\ll 1$. 
Thus, we need a high-sensitivity apparatus to make a precise measurement of $G$.  
Another reason comes from the universal nature of gravity; All particles or massive objects feel and produce gravitational forces. 
As a result, a small distortion of the experimental apparatus (such as external source) from its ideal shape directly produces a small change of gravitational force, and it causes a systematic error of the experiment.  
Therefore, we must prepare the very elaborate apparatus for the precise determination of $G$.  

In recent years, in addition to the traditional Cavendish-type experiments \cite{Quinn:2013pea} which are based on the torsion balance condition, conceptually different experiments are also performed.  
For example, in \cite{Parks:2010sk}, the authors determined $G$ by measuring the change of the length of Fabry-P{\'e}rot resonator caused by the external gravity source.   
In \cite{Rosi:2014kva}, $G$ is interferometrically determined by using many cold ${}^{87}$Rb atoms $(N\sim 10^{9})$ where excellent optical techniques such as the Raman transition method  are used.  
%
These conceptually different experiments are important in the sense that they help to identify new systematic errors which are difficult to capture in the traditional methods. 

In this letter, we investigate a new experimental possibility of measuring $G$ by using the cold atoms via the weak measurement \cite{Aharonov_1964,Aharonov:1988xu}.  
We study the motion of atoms which are gravitationally interacting with an external gravity source, and consider its weak measurement. 
Then, it is shown that such a motion can be amplified by the amount of $\mathcal{O}(10^{3})$ compared with their classical motion  
when the number of atoms is $\sim 10^{15}$ and the observing time is $\sim 0.5$s. 
This result may open a new possibility of determining $G$ based on the weak measurement. 
%
%
%
In Supplementary Material, we give a brief review of weak measurement for readers who are not familiar with this topic. 
%
%
\\ 
%
%
%

{\ \ {\it Cold atoms with external gravity source.---}}
%
\begin{figure}
\begin{center}
\includegraphics[width=.40\textwidth]{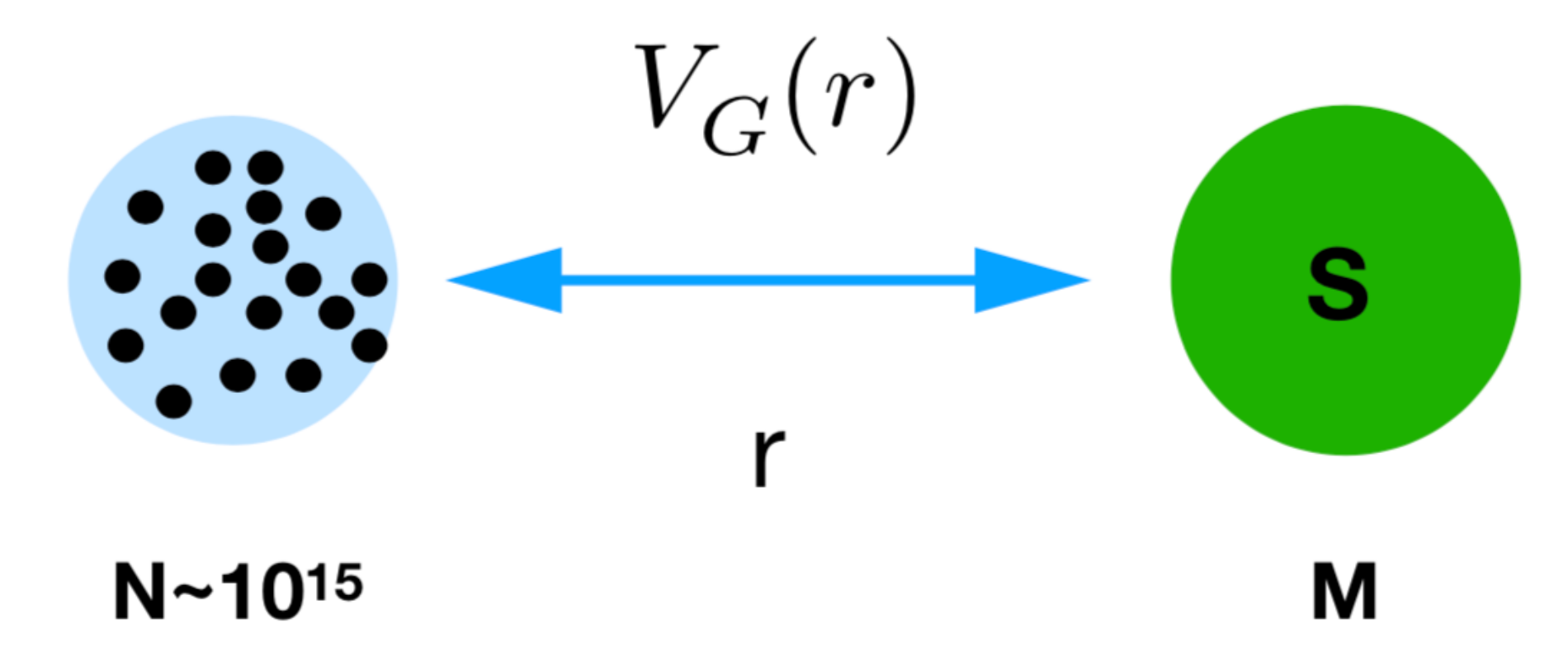}
\caption{A schematic figure of our set up. 
Here we show the case where the gravity source can be regarded as a massive particle.  
%
}
\label{fig:system}
\end{center}
\end{figure}
We want to consider the weak measurement of a cold atomic system which is gravitationally interacting with external gravity source whose mass is denoted by $M$. In Fig.\ref{fig:system}, we schematically show this set up. 
In particular, we consider the case where the relative angular momentum between atoms and external source is zero. 
Then, the Hamiltonian of this system is 
\be \hat{H}=\hat{H}_\text{atom}^{}+\hat{m}^{-1}\otimes \frac{\hat{p}_r^2}{2} +\hat{V}_G^{}(\hat{r}),\
\label{eq: total hamiltonian}
\e
where $\hat{H}_\text{atom}^{}$ is the Hamiltonian of atoms, $\hat{m}\ (\hat{m}^{-1})$ is the (inverse) mass operator of atoms including its binding energy which is defined below, $\hat{r}$ is the operator corresponding to relative distance between atoms and external source, and $\hat{V}_G^{}(\hat{r})$ represents the operator corresponding to their classical gravitational potential $V_G(r)$ whose functional form depends on the shape of the external gravity source. 
For example, if the external source can be regarded as a massive particle, 
 it becomes $ V_G^{}(r)=-GMmr^{-1}. 
\label{eq: r inverse}$ 
Or, if the external source is a cylinder with its mass density per unit length being $\rho$ and its radius being $l$, we have $V_G^{}(r) =2mG\rho\log\left(\frac{r}{le^{1/2}}\right),\  (\text{for }r\geq l)$  
where $e$ is the Napier's constant. 
%
%
In the following, we proceed with our argument without specifying a specific form of $\hat{V}_G^{}$ until we need to estimate physical quantities numerically. 
%
In Eq.(\ref{eq: total hamiltonian}), we have neglected the Hamiltonian corresponding to the each motion of atoms including the center-of-mass motion of the total system because it is irrelevant in the following theoretical discussion. 
\footnote{
In a realistic BEC such as Alkali atoms, they form a stretched cloud described by a macroscopic wave function \cite{BEC}, and we must consider this stretched effect when we observe these atoms. 
} 
Besides, we assume that the external gravity force 
is perpendicular to that of the earth so that we do not need to consider the latter effects.    
Then, the Hilbert space of this system is $ {\cal{H}}={\cal{H}}_A^{}\otimes {\cal{H}}_R^{}$ where ${\cal{H}}_A^{}\ ({\cal{H}}_R^{})$ is the Hilbert space of atom (the relative motion). 
From the point of view of quantum measurement, the atomic system corresponds to a measured system, and the relative motion corresponds to a probe system. 

In order to describe the atomic state, we follow the second quantization picture:
$ \hat{H}_\text{atom}^{}=\sum_{n=1}^{\infty}E_n^{}\hat{N}_n^{},\ 
\hat{m}=\sum_{n=1}^{\infty} m_n^{}\hat{N}_n^{}
$,  
where $\hat{N}_n^{}=\hat{a}_n^\dagger\hat{a}_n^{}$ is the number operator of the $n$-th eigenstate,
    and $E_n^{}$ is the energy eigenvalue of atom which is in principle determined by solving the Shr$\ddot{\text{o}}$dinger equation: $\left(-\hbar^2\nabla^2/2m_e^{}+V_\text{eff}^{}(\mathbf{x})\right)\psi_n^{}(\mathbf{x})=E_n^{}\psi_n^{}(\mathbf{x}), 
   $ 
   where $V_{\text{eff}}^{}(\mathbf{x})$ is the effective potential of an electron around a nucleus. 
For example, $ |1\rangle=\hat{a}_1^{\dagger}|0\rangle$ and $|2\rangle=\hat{a}_2^{\dagger}|0\rangle$ correspond to the ground state and 
the first excited state respectively, and their difference $(\Delta E=\Delta m=E_2^{}-E_1^{}\sim {\cal{O}}(10^{-5})\text{eV})$ typically originates in the hyper fine splitting. 
In $\mathcal{H}_A$, the basis vector can be expressed as $|l_1,l_2,\cdots\rangle\propto (\hat{a}_{1}^{\dagger})^{l_1}(\hat{a}_{2}^{\dagger})^{l_2}\cdots|0\rangle$  
where $l_n^{}$ corresponds to the number of particle in n-th state. 
In particular, the operation of $\hat{m}^{-1}$ is defined as $\hat{m}^{-1}|l_1,l_2,\cdots\rangle=(\sum_{n=1}^{\infty}m_n l_n )^{-1}|l_1,l_2,\cdots\rangle$.   
Note that we can neglect the effect of atomic transition between these two states because its time-scale is much larger than that of the observing time $\sim 0.5{\rm s}$. 
   A large number of atoms which are in a state of Bose-Einstein condensate are described by the coherent state $|n;N_n^{}\rangle$ which is defined as $|n;N_n^{}\rangle=e^{-N_n /2}e^{\sqrt{N_n}\hat{a}^{\dagger}_n}|0\rangle$ and satisfies $\hat{a}_n^{}|n;N_n^{}\rangle=\sqrt{N_n^{}}|n;N_n^{}\rangle$ where $N_n^{}$ is the number of $n$-th eigenstate atoms. 
%
\footnote{Considering such a coherent state is not crucial in the following discussion. 
We can also obtain the same result even if the state is not the coherent one such as 
\be \frac{1}{\sqrt{N!}}\hat{a}_i^{\dagger N}|0\rangle.
\e
}
In the following, it is sufficient to consider the restricted Hilbert space spanned by $|1;N_1^{}\rangle$ and $|2;N_2^{}\rangle$  
because we are considering the transition between these two states. 
%

The analytical treatment of this system is not so easy unless we make a few approximations: 
%
%
\begin{enumerate}
\item We define a new coordinate $x$ instead of $r$: $ r=R+x$  
where $R$ is the initial distance between atoms and external source. 
Then, we consider the following potential:
\aln{V_G^{}(r)=V_G^{}(R)+\frac{dV_G^{}(R)}{dr}x+{\cal{O}}(x^2). 
\label{eq: potential expansion}
}
which corresponds to the leading order expansion. 
We expect that the dynamics of the present system is well described by Eq.(\ref{eq: potential expansion}) as long as $ x\ll R$.  
The Hamiltonian is now approximated as $
\hat{H}
\equiv \hat{H}_0^{}+\hat{H}_1^{}$ 
where
\small
\be \hat{H}_0^{}=\hat{H}_\text{atom}^{}+\hat{V}_G^{}(R),\ \hat{H}_1^{}=\hat{m}^{-1}\otimes\frac{\hat{p}^2}{2}+\frac{d\hat{V}_G^{}(R)}{dr}\hat{x},
\e
\normalsize
and they satisfy $[\hat{H}_0^{},\hat{H}_1^{}]=0 \label{eq:commutation}$. 
Here, we defined $\hat{V}_G(R)$ and $d\hat{V}_G(R)/dr$ as operators corresponding to $V_G(R)$ and $d{V}_G(R)/dr$. 
For example, in the case of the $r^{-1}$ potential, they are $\hat{V}_G(R)\equiv -GMR^{-1}\hat{m}$ and $d\hat{V}_G(R)/dr\equiv GMR^{-2}\hat{m}$  respectively. 
%
%
\item As for the initial state of the relative motion, we assume a gaussian wave packet: ${\langle x|\phi_i\rangle^{}}=e^{-\frac{x^2}{2d^2}}/(\pi d^2)^{1/4},\label{eq: initial wave}$ 
where the width $d$ also depends on details of experimental set up. 
For example, in the atomic interferometer experiment \cite{Rosi:2014kva}, $d$ is ${\cal{O}}(1\text{mm})$. 
So we choose $d=1$mm as a typical value for our estimation.  
\end{enumerate}
Based on these approximations, we can actually perform analytical calculation. 
In particular, the time evolution can be completely solvable without any further approximations. 
However, because the details of such calculations are rather cumbersome and physically not so clear, we will give a leading order calculation in the following. \\  

{{\it Weak measurement.---}}
Let us now consider the weak measurement of this atomic system. 
What we want to know is how large the expectation value of $\hat{x}$ can be amplified compared with that of the classical motion, i.e. 
$ x_{\text{cl}}^{}(t)\simeq -\frac{t^2}{2m}\frac{dV_G^{}(R)}{dr}$. 
%
In the following, we give the leading order calculation with respect to $G$. 
See Supplementary Material for the detailed all-order calculations. 

\begin{figure}[t]
\begin{center}
\includegraphics[width=.40\textwidth]{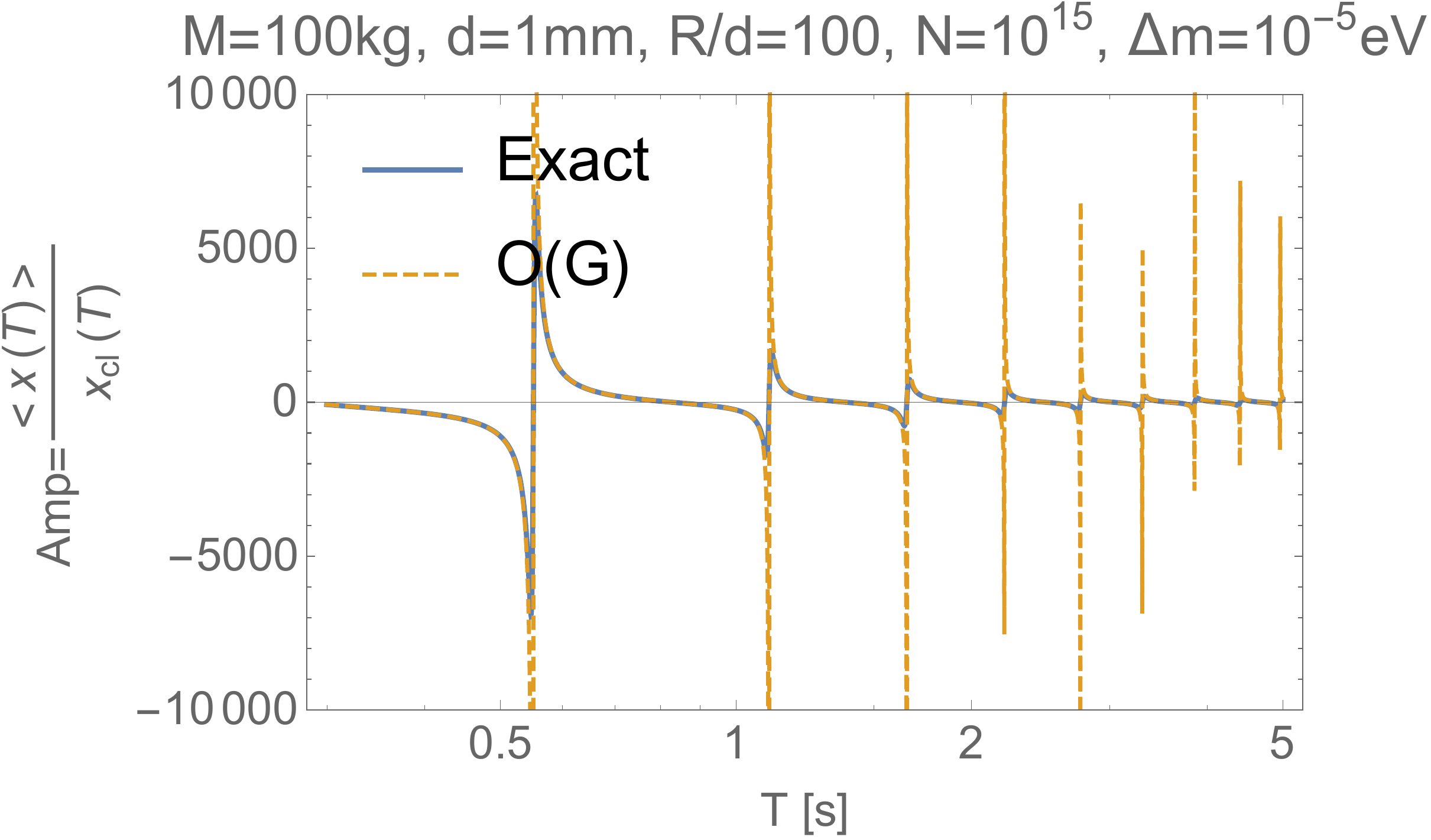}
\includegraphics[width=.40\textwidth]{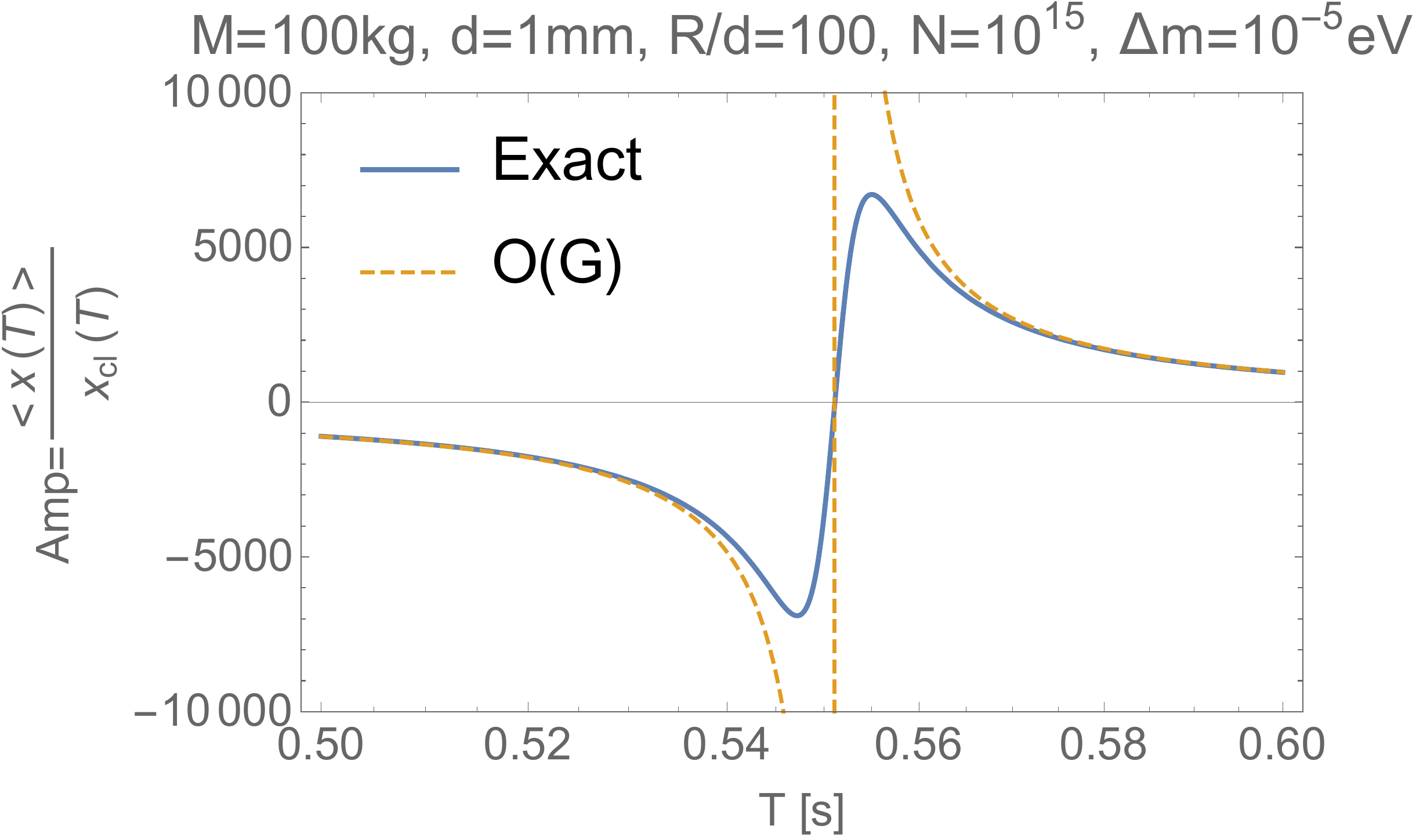}
\caption{Plots of the amplification factor as a function of the observing time $T$. 
Here the dashed orange lines correspond to the leading order result  Eq.(\ref{eq: leading result}), and the blue lines represent the exact results Eq.(\ref{eq: exact result}).  
The lower figure is the enlarged plot of the upper figure around the first peak.
}
\label{fig:plot}
\end{center}
\end{figure}

We can evaluate the time evolution as 
\small
\aln{
e^{-i\hat{H}_1^{}t}e^{-i\hat{H}_0^{}t} |\psi_i\rangle^{}|\phi_i\rangle^{}
\simeq &
e^{-i\frac{d\hat{V}_G^{}(R)}{dr}\hat{x}t}e^{-i\hat{H}_0^{}t} |\psi_i\rangle^{}e^{-i\frac{\hat{p}^2}{2\overline{m}}t-ix_{\text{cl}}^{}(t)\hat{p}} |\phi_i\rangle^{}
\nn
\equiv & e^{-i\frac{d\hat{V}_G^{}(R)}{dr}\hat{x}t} |\psi(t)\rangle^{}|\phi(t)\rangle^{},
}
\normalsize
where $  |\psi(t)\rangle^{}=e^{-i\hat{H}_0^{}t} |\psi_i\rangle^{},\ 
  |\phi(t)\rangle^{}=e^{-i\frac{\hat{p}^2}{2\overline{m}}t-ix_{\text{cl}}^{}(t)\hat{p}} |\phi_i\rangle^{}$. 
Here, we have assumed that the mass operator in the kinetic term were constant $\bar{m}$ for simplicity. 
Then, after the post-selection of atoms at $t=T$, the probe's wave function 
 is given by
\aln{  |\phi_f(T)\rangle^{}&\equiv {}^{}\langle \psi_f|e^{-i\frac{d\hat{V}_G^{}(R)}{dr}\hat{x}T}|\psi(T)\rangle^{}|\phi(T)\rangle^{}
\nn
\simeq &  {}^{}\langle \psi_f|\psi(T)\rangle^{}\exp\left(-iV_G^W
\hat{x}\right) |\phi(T)\rangle^{},
\label{eq: probe wave}
}
where
\be V_G^W
\equiv \frac{ \langle \psi_f|\frac{d\hat{V}_G^{}(R)}{dr}|\psi(T)\rangle^{}}{{}^{}\langle \psi_f|\psi(T)\rangle^{}}\times T
\label{eq: weak value of gravity}
\e
is the weak value in this system. 
%
Here, note that the normalization of $ |\phi_f(T)\rangle^{}$, i.e. $ P_{\text{tran}}^{}(T)\equiv  {}^{}\langle\phi_f(T)|\phi_f(T)\rangle^{}$ represents the transition (conditional) probability with which the process $|\psi_i\rangle^{}\rightarrow|\psi_f\rangle^{}$ occurs.     
%
%
Because $\hat{x}$ exists in the exponent in Eq.(\ref{eq: probe wave}),   
the shift of the peak of $ |\phi(T)\rangle^{}$ is now caused by the imaginary part of the weak value: 
%
\be  \frac{^{}\langle \phi_f(T)|\hat{x}|\phi_f(T)\rangle^{}}{^{}\langle \phi_f(T)|\phi_f(T)\rangle^{}}\simeq x_{\text{cl}}^{}(T)+ d^2 \times \text{Im}\left(V_G^W\right).\label{eq: leading x}
\e 
For example, in the case of the $r^{-1}$ potential
, if we choose $  |\psi_i\rangle^{}=\frac{1}{\sqrt{2}}(|1;N\rangle+|2;N\rangle),\  |\psi_f\rangle^{}=\frac{1}{\sqrt{2}}(|1;N\rangle-|2;N\rangle)
\label{eq: initial and final}
$
as the initial and final states \footnote{\label{foot:sel}
These two states are the eigenstates of $\hat{V}_{E}^{}=-\hat{\mathbf{q}}\cdot \mathbf{E}(t),$ where $\hat{\mathbf{q}}$ is the electric dipole moment of an atom and $\mathbf{E}(t)$ is the external electric field.  
Thus, we can choose these states for the pre- and post-selected states as in the case of the Stern-Gerlach experiment.  
}
, the amplification factor becomes 
%
\footnotesize 
\aln{&\text{Amp}\equiv {\frac{1}{x_{\text{cl}}^{}(T)}\frac{^{}\langle \phi_f(T)|\hat{x}|\phi_f(T)\rangle^{}}{\langle \phi_f(T)|\phi_f(T)\rangle^{}}=1+\frac{N\Delta md^2}{T} \frac{\sin f(T)}{1-\cos f(T)}}
\nn
&{\sim 1+10^3\left(\frac{N}{10^{15}}\right)\left(\frac{\Delta m}{10^{-5}\text{eV}}\right)\left(\frac{d}{1\text{mm}}\right)^2\left(\frac{0.5\text{sec}}{T}\right)
 \frac{\sin f(T)}{1-\cos f(T)}},
\label{eq: leading result}
}
\normalsize
where $\Delta m=m_2^{}-m_1^{}(=E_2^{}-E_1^{})$ and
\small  
\aln{ f(T)&+N\Delta mT\equiv \frac{GM \Delta m NT}{R}
\nn
\sim &2\pi \left(\frac{N}{10^{15}}\right)\left(\frac{M}{100\text{kg}}\right)\left(\frac{\Delta m}{10^{-5}\text{eV}}\right)\left(\frac{T}{0.5\text{sec}}\right)\left(\frac{10\text{cm}}{R}\right).
\label{eq: phase function}
}
\normalsize
Here, compared with the exact result presented in Supplementary Material, we have no exponential damping factor in Eq.(\ref{eq: leading result}). 
%
%
In general, the above amplification factor oscillates violently as a function of the observing time $T$ because of the phase $N\Delta m T$ which originates in $\hat{H}_{\text{atom}}^{}$.
On the other hand, 
Eq.(\ref{eq: phase function}) has a longer time scale $\sim 1$s for the typical values of parameters. 
Thus, if the phase $N\Delta m T$ can be removed by some mechanism 
\footnote{
For example, if we can observe $T$ in units of $2\pi/\Delta m\sim 10^{-10}$s 
 this phase has no effects. 
We hope that this kind of observation can be realized by using modern technology of optics, but discussing it is beyond the scope of this letter. 
%
%
%
%
%
}, 
it is possible to obtain a large amplification of the relative motion around $T\sim 0.5$s.
%
In Fig.\ref{fig:plot}, we plot Eq.(\ref{eq: leading result}) (dashed orange line) along with the exact analytical result Eq.(\ref{eq: exact result}) 
(blue line) presented in Supplementary Material where the phase $N\Delta m t$ is omitted. 
As for the leading order result, the expectation value diverges when $f(T)=2n\pi\ (n=1,2,\cdots)$. 
On the other hand, the exact result has finite peaks, and their positions are slightly different from that obtained by  $f(T)=2n\pi\ (n=1,2,\cdots)$. 
More generally, in \cite{Lee Tsutsui}, it was shown that the amplification of observable is constrained by the effects of the higher order corrections and the systematic and statistical uncertainties. 
Although we do not consider the latter effects here, our result is an explicit example of this general argument.   
%
This fact means that 
$P_{\text{tran}}^{}(T)$ is also finite at these peaks because of the effects of the higher order corrections.       
By using Eq.(\ref{eq: exact result}), it can be roughly estimated as
$
P_{\text{tran}}^{}(T)\sim \frac{\pi^2}{2}\left(d/R\right)^2\sim 3\times 10^{-4}\ (\text{for } d=1\text{mm},\ R=10\text{cm}),  
$
which leads to the number of surviving atoms as $\sim 10^{11}$.  
Fig.\ref{fig:plot} also shows that the amplification can be negative, and this means that the gravitational force can behave like a repulsive force. 
Qualitatively, this negative amplification comes from the fact that the weak value Eq.(\ref{eq: weak value of gravity}) can generally take  both of negative and positive values. 
Although this behavior seems to be unnatural from the usual attractive property of gravity, 
our result shows that a very small number of atoms actually feel such a repulsive force due to the quantum effects.      
%
%
%

The weak measurement can also amplify the fluctuation of $\hat{x}$.    
If this is the case, combined with the small transition probability, the measurement of a large weak value becomes more and more difficult. 
In our case, however, this does not occur. 
In fact, at the leading order, we have 
\aln{ \Delta_x^{2}(T)\equiv  \frac{^{}\langle \phi_f(T)|(\hat{x}-\langle \hat{x}\rangle)^2|\phi_f(T)\rangle^{}}{^{}\langle \phi_f(T)|\phi_f(T)\rangle^{}}
=d(T)^2+{\cal{O}}(G^2),
}
and this result comes from the fact that the initial wave function Eq.(\ref{eq: initial wave}) is invariant under $x\rightarrow -x$. 
In other word, 
in the case of a distorted initial atomic cloud,  
the fluctuation of the relative position is also amplified by the weak measurement. 
We have also numerically checked that this conclusion does not much change even if we take the higher order corrections into account. 
%
%
%
\\ 
%
%
%

{\it Conclusion.---}
In this letter, we have considered the weak measurement of a cold  atomic system which is gravitationally interacting with an external gravity source.  
From the point of view of quantum measurement, their relative motion can be naturally regarded as a probe system which can be used for measuring the atomic system indirectly.  
Then, we have shown that it is actually possible to realize ${\cal{O}}(10^{3})$ amplification compared with the classical motion when $N\sim 10^{15}$.  
In particular, we have seen that the peak of the amplification is  finite due to the effects of the higher order contributions.  
This result means that the ordinary argument based on the leading order calculation is not quantitatively correct. 
Thus, when one wants to consider an application of weak measurement to some quantum interacting system, it is necessary to take higher order corrections into account in order to obtain quantitatively correct predictions. 
Then, we have also seen that the gravitational force can behave as a repulsive force in response to a negative weak value. 
This counterintuitive result is also one of the interesting aspects of weak measurement. 
%

%
%
%
%

%
Although experiential realization of our set up seems to be difficult for the present time, we hope that this kind of experiment will be performed in the near future. 
Besides, we expect that our work stimulates the discussion of weak measurement and its application to gravity. 
%

\subsection*{Acknowledgement}
%
We would like to thank Motoi Endo, Satoshi Iso, Hikaru Kawai, Izumi Tsutsui, Ryuichiro Kitano, Masaki Ando, Ryota Kojima, Jaeha Lee, Yuichiro Mori, Yosuke Morimoto, Katumasa Nakayama, Hikaru Ohta, Sayuri Takatori, Yoshinori Tomiyoshi and Sumito Yokoo for helpful discussions. 
The work of KK is supported by the Grant-in-Aid for JSPS Research Fellow, Grant Number 17J03848.


%
%
%
%
\clearpage 
\begin{widetext}
\section*{Supplementary Material for “Amplification of gravitational motion via Quantum weak measurement”}
%
%
This supplemental material provides (I) the brief review of weak measurement and (II) the full order calculation of the expectation value of $\hat{x}$.  \\
\\
(I){\it Brief review of weak measurement.---}
The idea of weak measurement or weak value was originally proposed in \cite{Aharonov_1964,Aharonov:1988xu} 
where the authors considered the Stern-Gerlach experiment and showed that the measured value of electron's spin can become quite large $\gg 1/2$ by choosing its initial and final states artificially so that they are nearly orthogonal each other. 
This kind of amplification of observables is a general feature of weak measurement, and its qualitative understanding can be easily grasped by considering a von-Neumann type system \cite{von Neumann}.   
Suppose that we are considering an interacting system $A\otimes B$ where $A$ is some quantum system which we want to measure and $B$ is a probe system having a coordinate degree of freedom $\hat{x}$.  
For example, in the original paper \cite{Aharonov:1988xu}, $A$ is the spin of electron and $B$ is its position $\hat{z}$.  
Then, as the Hamiltonian of this system, we consider 
\aln{ \hat{H}_{VN}^{}=g\hat{{\cal{O}}}_A^{}\otimes\hat{p}\times \delta(t-t_0^{})\ (t_0^{}>0),
} 
where $\hat{{\cal{O}}}_A^{}$ is an observable of $A$, $\hat{p}$ is the conjugate momentum of $\hat{x}$ and $g\ll 1$ is a {\it weak}  coupling constant.
The time evolution can be solved as 
\aln{|\Psi\rangle^{}\equiv e^{-i\int_0^t dt'\hat{H}_{VN}^{}}|\psi_i\rangle^{}\otimes |\phi_i\rangle^{}=e^{-ig\hat{{\cal{O}}}_A^{}\hat{p}}|\psi_i\rangle^{}\otimes |\phi_i\rangle^{}, 
} 
where $|\psi_i\rangle^{}\ (|\phi_i\rangle^{})$ is an initial state of $A\ (B)$. 
In the following, we assume that $|\phi_i\rangle^{}$ has a unique peak at $x=0$ for simplicity. 
In the ordinary measurement, we measure the distribution of $x$, i.e. 
\aln{ P(x)=\langle x|\text{Tr}_A^{}(|\Psi\rangle^{}{}^{}\langle\Psi|)|x\rangle
=\sum_k |\langle a_k^{}|\psi_i\rangle^{}|^2\times |\langle x|e^{-iga_i^{}\hat{p}}|\phi_i\rangle^{}|^2,
\label{eq: ordinary distribution}
}
where $|a_k^{}\rangle\ (a_k^{})$ is the eigenstate (eigenvalue) of $\hat{{\cal{O}}}_A^{}$. 
Here, the wave function $|\langle x|e^{-iga_k^{}\hat{p}}|\phi_i\rangle^{}|^2$ has a peak at $x=ga_k^{}$ because $e^{-iga_k^{}\hat{p}}$ acts as a translation operator.  
Therefore, $P(x)$ typically has a shape like Fig.\ref{fig:ordinary} where each of the peaks corresponds to the eigenvalue $a_k^{}$. 
Thus, by observing $P(x)$, we can obtain the information of $a_i^{}$ and the expansion coefficient $|\langle a_k^{}|\psi_i\rangle^{}|$.  
%
\begin{figure}
\begin{center}
\includegraphics[width=.50\textwidth]{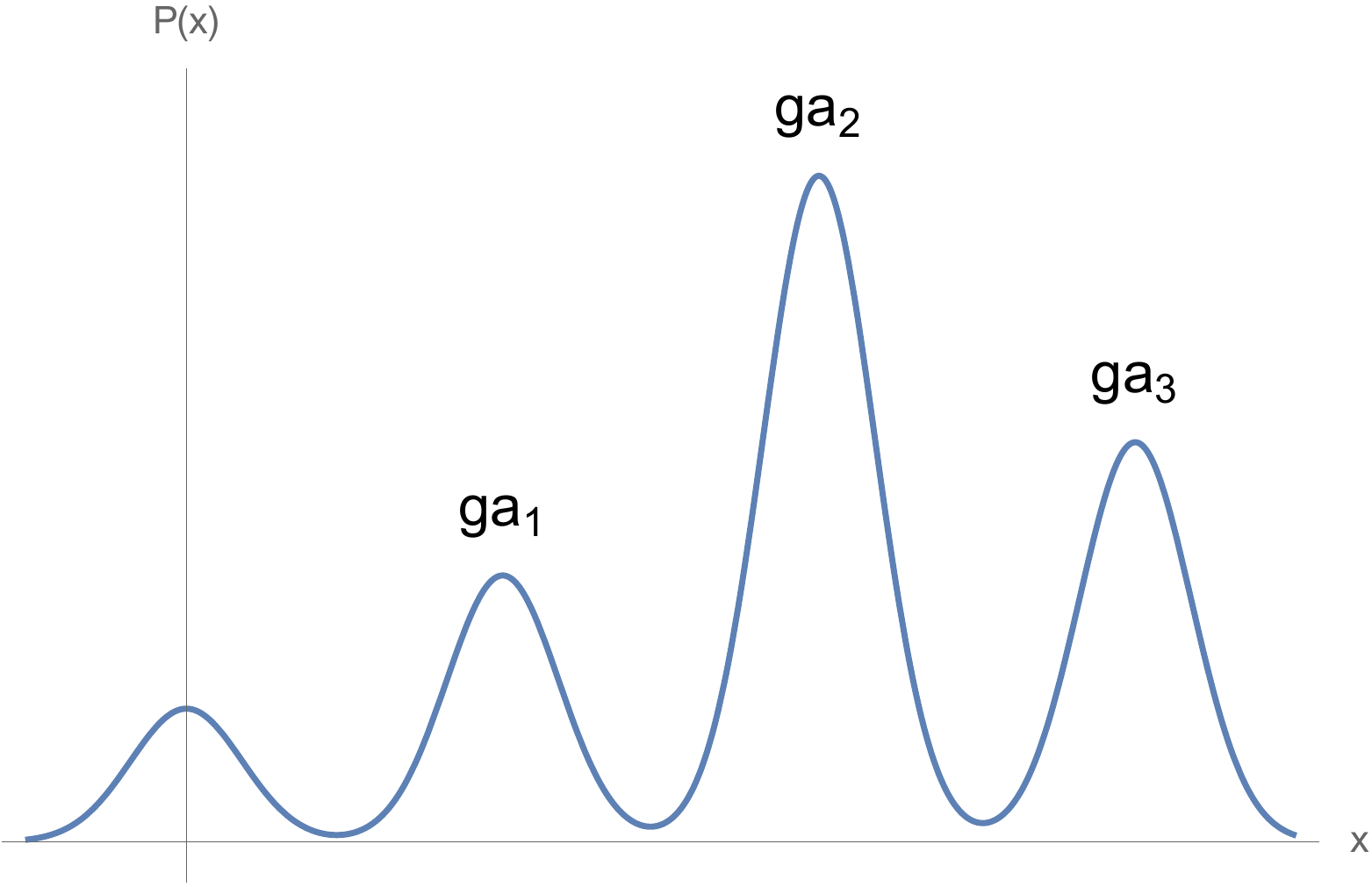}
\caption{A typical provability distribution of the probe's wave function in the ordinary indirect measurement.
Each of the peaks corresponds to the eigenvalue of ${\cal{O}}_A^{}$. 
}
\label{fig:ordinary}
\end{center}
\end{figure}

Now let us consider the weak measurement.
Suppose that we have prepared an experimental apparatus such that it enables us to restrict the final state of $A$ to a specific state $|\psi_f\rangle^{}$. (Post-selection) 
\footnote{
In order to do such a selection, we usually need another system such as external electromagnetic field as in the case of Stern-Gerlach experiment.  
Strictly speaking, we must also consider the effects of this external field for completely understating the process of the weak measurement.  
In this letter, however, we just concentrate on what happens by 
assuming that such a selection is realized.   
See also footnote \ref{foot:sel}. 
}
Then, the wave function of B under such restriction is given by 
\aln{|\phi_f\rangle^{}\equiv {}^{}\langle \psi_f|e^{-ig\hat{{\cal{O}}}_A^{}\hat{p}}|\psi_i\rangle^{}|\phi_i\rangle^{}
\sim  {}^{}\langle \psi_f|\psi_i\rangle^{}e^{-ig{\cal{O}}_A^W\times \hat{p}}|\phi_i\rangle^{},\label{eq: typical leading order calculation}
} 
where 
\aln{ {\cal{O}}_A^W\equiv {}^{}\langle \psi_f|\hat{{\cal{O}}}_A^{}|\psi_i\rangle^{}/{}^{}\langle \psi_f|\psi_i\rangle^{}
\label{eq: weak value 1}
} 
is called {\it the weak value} of $\hat{\cal{O}}_A^{}$. 
Here, compared to the wave function of the ordinary measurement, we have the weak value in the exponent of the translation operator. 
%
As a result, we have a unique peak at $g\text{Re}({\cal{O}}_A^W)$, and it can become larger than $g\underset{k}{\text{Max}}(a_k^{})$ if we choose the  initial and final states so that they satisfy ${}^{}\langle \psi_f|\psi_i\rangle^{}\sim 0$. 
In this sense, the weak measurement can amplify the measured values of observables. 
A few comments are needed here:
\begin{itemize}
\item Although the weak value ${\cal{O}}_A^W$ is large when ${}^{}\langle \psi_f^{}|\psi_i^{}\rangle\sim 0$, this means that we have little chance to observe such transition $|\psi_i\rangle^{}\rightarrow|\psi_f\rangle^{}$. 
Thus, if one wants to obtain a large weak value, it is necessary to consider a good experiment such that we can easily obtain large statistics.   
In our case, we expect that it is realized by preparing a large number of atoms. 
%
This qualitative argument can quantitatively change by the effects of the higher order corrections. 
In particular, 
the transition probability can be finite even at the peak of the weak value. 
The importance of this behavior was discussed in 
\cite{Lee Tsutsui} where the authors also considered the effects of systematic and statistical errors. 
\item In general, the weak measurement can also amplify the fluctuation  
\be \Delta_x^2\equiv \frac{{}^{}\langle \phi_f|\hat{x}^2|\phi_f\rangle^{}}{{}^{}\langle \phi_f|\phi_f\rangle^{}}-\left(\frac{{}^{}\langle \phi_f|\hat{x}|\phi_f\rangle^{}}{{}^{}\langle \phi_f|\phi_f\rangle^{}}\right)^2 
\e 
as well as the expectation value of $\hat{x}$. 
If this is the case, combined with the small transition probability, the measurement of a large weak value becomes more and more difficult. 
In our set up, however, it does not happen as we will see in the following.  
%
%
\end{itemize}
(II){ \it {Full order calculations.---}} Here, we present the full order calculations of the expectation value of $\hat{x}$. 
The probe's wave function after the post-selection is
\eqn{
|\phi_f^{}(T)\rangle &= &
\langle\psi_f^{}|\exp\left(-i\hat{H}_1^{}T\right)\exp\left(-i\hat{H}_0^{}T\right)|\psi_i^{}\rangle|\phi_i^{}\rangle
\nn
&=&\sum_{j=1}^2\langle\psi_f^{}|j;N\rangle
\langle j;N|\psi_i^{}\rangle
\exp\left(-i
\langle \hat{V}_G^{}(R)\rangle T-iNE_j^{}T\right)\exp\left(-i\frac{\hat{p}^2}{2Nm_i^{}}T
-iT\left\langle\frac{d\hat{V}_G^{}(R)}{dr}\right\rangle_j^{}
\hat{x}\right)|\phi_i^{}\rangle
\nn
&\equiv & \sum_{j=1}^2 c_i^{}
\exp\left(-i\frac{\hat{p}^2}{2Nm_i^{}}T
-iT\left\langle\frac{d\hat{V}_G^{}(R)}{dr}\right\rangle_j^{}
\hat{x}\right)|\phi_i^{}\rangle
\nn
&=&\sum_{j=1}^2 c_j^{}\exp\left(-i\frac{2Nm_j^{}}{3T}x_{\text{cl}}(T)^2\right)
\exp\left(-i\frac{T}{2Nm_j^{}}\hat{p}^2+ix_{\text{cl}}^{}(T)\hat{p}\right)
\exp\left(i\frac{2Nm_j^{}}{T}x_{\text{cl}}^{}(T)\hat{x}\right)|\phi_i^{}\rangle,
\label{eq: exact wave function}
}
where
\eqn{
c_j^{}&=&\langle\psi_f^{}|j;N\rangle
\langle j;N|\psi_i^{}\rangle
\exp\left(-i
\langle \hat{V}_G^{}(R)\rangle_j^{} T-iNE_j^{}T\right),
}
and we have used the Campbell-Baker-Hausdorff formula:
\be e^{A}e^{B}=e^{A+B+\frac{1}{2}[A,B]+\frac{1}{12}[A-B,[A,B]]+\cdots }.
\e
By inserting the complete set
\be 1=\int dp|p\rangle \langle p|\text{ or }\int dx|x\rangle\langle x|, 
\e
and performing the space and momentum integrations, we obtain 
\aln{
\langle \phi_f^{}(T)|\phi_f^{}(T)\rangle\simeq |c_1^{}|^2+|c_2^{}|^2
+\exp \left(-\frac{N^2d^2\Delta m^2}{T^2}x_{\text{cl}}^{}(T)^2\right) 
\times \left(c_1^*c_2^{}\exp\left(-i\frac{2N\Delta m}{3T}x_{\text{cl}}^{}(T)^2
\right)+\text{h.c}
\right),
}
\aln{ \langle \phi_f^{}(T)|\hat{x}|\phi_f^{}(T)\rangle &\simeq 
(|c_1^{}|^2+|c_2^{}|^2)x_{\text{cl}}^{}(T)
+\frac{1}{2}\exp \left(-\frac{N^2d^2\Delta m^2}{T^2}x_{\text{cl}}^{}(T)^2\right) 
\nn
&
\times \left[c_1^*c_2^{}\exp\left(-i\frac{2N\Delta m}{3T}x_{\text{cl}}^{}(T)^2\right)
\times \left(-2x_{\text{cl}}^{}(T)+2i\frac{d^2N\Delta m}{T}x_{\text{cl}}^{}(T)+\left(2+\frac{m_2^{}}{m_1^{}}+\frac{m_1^{}}{m_2^{}}\right)x_{\text{cl}}^{}(T)
\right)
+\text{h.c}
\right],
}
where we have neglected the terms containing 
\be \frac{2T\Delta m}{N m_1^{}m_0^{}}\ll 1.
\e
Then, if we choose Eq.(\ref{eq: initial and final}) as the initial and final states, we obtain 
\eqn{ \frac{\langle \phi_f^{}(T)|\hat{x}|\phi_f^{}(T)\rangle}{\langle \phi_f^{}(T)|\phi_f^{}(T)\rangle}\bigg/x_{\text{cl}}(T)\simeq 1-\frac{Nd^2 \Delta m}{T}\frac{e^{-g(T)}\sin f(T)}{1-e^{-g(T)}\cos f(T)},
\label{eq: exact result}
}
where
\aln{ &g(T)=\frac{N^2d^2\Delta m^2}{T^2}x_{\text{cl}}^{}(T)^2,
\ f(T)=\frac{2N\Delta m}{3T}x_{\text{cl}}^{}(T)^2+T\left(\langle \hat{V}_G^{}(R)\rangle_2^{}-\langle \hat{V}_G^{}(R)\rangle_1^{}\right)+TN\Delta m.
}
\end{widetext}


\begin{thebibliography}{99}

\bibitem{Mohr:2015ccw} 
  P.~J.~Mohr, D.~B.~Newell and B.~N.~Taylor,
  ``CODATA Recommended Values of the Fundamental Physical Constants: 2014,''
  Rev.\ Mod.\ Phys.\  {\bf 88}, no. 3, 035009 (2016)
  doi:10.1103/RevModPhys.88.035009
  [arXiv:1507.07956 [physics.atom-ph]].

\bibitem{Quinn:2013pea} 
  T.~Quinn, H.~Parks, C.~Speake and R.~Davis,
  ``Improved Determination of G Using Two Methods,''
  Phys.\ Rev.\ Lett.\  {\bf 111}, no. 10, 101102 (2013).
  doi:10.1103/PhysRevLett.113.039901, 10.1103/PhysRevLett.111.101102

\bibitem{Parks:2010sk} 
  H.~V.~Parks and J.~E.~Faller,
  ``A Simple Pendulum Determination of the Gravitational Constant,''
  Phys.\ Rev.\ Lett.\  {\bf 105}, 110801 (2010)
  doi:10.1103/PhysRevLett.105.110801
  [arXiv:1008.3203 [physics.class-ph]].

\bibitem{Rosi:2014kva} 
  G.~Rosi, F.~Sorrentino, L.~Cacciapuoti, M.~Prevedelli and G.~M.~Tino,
  ``Precision Measurement of the Newtonian Gravitational Constant Using Cold Atoms,''
  Nature {\bf 510}, 518 (2014)
  doi:10.1038/nature13433
  [arXiv:1412.7954 [physics.atom-ph]].
  
 \bibitem{Aharonov_1964}
Y. Aharonov, P. G. Bergmann and L. Lebowitz,
``Time Symmetry in the Quantum Process of Measurement,''
Phys.\ Rev.\ {\bf 134} (1964) B1410. 
doi:10.1103/PhysRev.134.B1410
    
\bibitem{Aharonov:1988xu} 
  Y.~Aharonov, D.~Z.~Albert and L.~Vaidman,
  ``How the result of a measurement of a component of the spin of a spin-1/2 particle can turn out to be 100,''
  Phys.\ Rev.\ Lett.\  {\bf 60}, 1351 (1988).
  doi:10.1103/PhysRevLett.60.1351

\bibitem{von Neumann}   
  J. von Neumann, Mathematische Grundlagen der Quan temechanik (Springer-Verlag, Berlin, 1932) [English transla- tion: Mathematical Foundations of Quantum Mechanics (Princeton Univ. Press, Princeton, N3, 1955)l.
 
\bibitem{Lee Tsutsui}   
  Jaeha Lee and Izumi Tsutsui,  ``Merit of amplification by weak measurement in view of measurement uncertainty,''
  Quantum Stud.: Math. Found. (2014) 1: 65.
 
 \bibitem{BEC}
 F. Dalfovo, S. Giorgini, L. P. Pitaevskii, and S. Stringari, ``Theory of Bose-Einstein condensation in trapped gases,'' Rev. Mod. Phys. 71, 463 (1999).
  
  
 
\end{thebibliography}
\end{document}